\documentstyle[preprint,aps,psfig,epsfig]{revtex}

\begin{document}

\author{A. Perali, P. Pieri, and G. C. Strinati}

\address{Dipartimento di Fisica,
Sezione INFM, Universit\`a di Camerino,\\
Via Madonna delle Carceri, I-62032 - Camerino, Italy}

\date{\today}

\title{Comment on ``BCS to Bose-Einstein crossover phase diagram at zero
temperature for a $d_{x^2-y^2}$ order parameter superconductor: Dependence
on the tight-binding structure''}

\maketitle

\begin{abstract}
The work by Soares {\em et al.} [Phys. Rev. B {\bf 65}, 
174506 (2002)] investigates the BCS-BE crossover for $d$-wave pairing in
the 2-dimensional attractive Hubbard model. Contrary to their claims, 
we found that a non-pairing region does {\em not} exist in the
density vs coupling phase diagram. The gap parameter at $T=0$,
as obtained by solving analytically as well as numerically 
the BCS equations, is in fact finite for any non-zero density 
and coupling, even in the weak-coupling regime.
\end{abstract}

In Ref. \onlinecite{Soares}, Soares {\em et al.} analyzed the BCS-BE crossover
at $T=0$ for $d$-wave pairing in the attractive Hubbard model, 
extending the work by den Hertog \cite{Hertog} by
including the next-nearest neighbor hopping $t^\prime$ in the tight-binding
dispersion. As is evident from the insets of Figs. 1 and 3 as well as from
the phase diagram of Figs. 2 and 4 of Ref. \onlinecite{Soares}, in the 
weak-coupling regime and for any density, 
they obtained a non-pairing
metallic phase characterized by an exactly vanishing 
value of the BCS gap parameter
at $T=0$ $(\Delta (T=0) = 0)$. 
Similar results have been previously obtained by den Hertog 
in Ref. \onlinecite{Hertog} for $t^\prime=0$. 
In a related context,
a metallic phase at $T=0$ in the weak-coupling regime could only result
from the Capone {\em et al.} calculation in Ref. \onlinecite{Capone} 
(using a DMFT approach to the $s$-wave attractive Hubbard model at $T=0$) 
by suppressing superconducting pairing.
On the other hand, results for the gap parameter obtained with Hubbard-like
lattice models can be connected (in the weak-coupling and low-density 
limits) to the analytic results of the continuous models (see below) 
obtained in Refs. \onlinecite{Maki} and \onlinecite{Musaelian}, 
where a {\em finite} value for the gap is found in the weak-coupling regime.

In this Comment, we show that the $d$-wave 
gap parameter at $T=0$ is {\em always finite} for 
any non-zero density and even in the weak-coupling regime, by 
performing both numerical and analytic calculations.
This result was reported 
in Ref. \onlinecite{Andrenacci} where it was explicitly contrasted 
with the results by den Hertog \cite{Hertog}; 
further comments on this point can be 
found in the review article by Loktev {\em et al.} \cite{Loktev}. 
In particular, Soares {\em et al.} seem not to have been aware of the results 
by Ref. \onlinecite{Andrenacci} concerning the weak-coupling regime.
The difference in the results must be due to inaccuracy in the 
numerical calculations of Refs. \onlinecite{Soares} and \onlinecite{Hertog}.

To further support our previous finding about the existence of a $d$-wave 
BCS superconducting ground state ($\Delta (T=0) \neq 0$) in the 
weak-coupling regime, we solve the 
coupled self-consistent equations for the gap function and density 
at $T=0$, as given by BCS theory:
\begin{equation}
\label{bcsgap}
\Delta_{\bf k}=-\frac{1}{N}\sum_{\bf k^\prime}\frac{V_{{\bf k},
{\bf k^\prime}}}
{2\sqrt{\xi_{\bf k^\prime}^2+\Delta_{\bf k^\prime}^2}}\Delta_{\bf k^\prime}
\end{equation}
\begin{equation}
\label{bcsdens}
n=\frac{1}{N}\sum_{\bf k}\left(1-\frac{\xi_{\bf k}}
{\sqrt{\xi_{\bf k}^2+\Delta_{\bf k}^2}}\right).
\end{equation}
Here, $\xi_{\bf k}=-2t(\cos k_x+\cos k_y)-4t^\prime\cos k_x\cos k_y-\mu$
is the tight-binding dispersion with nearest- and next-nearest-neighbor
hopping, $\mu$ the chemical potential,
$N$ the total number of sites, and 
$V_{{\bf k},{\bf k^\prime}}=-V\gamma_{\bf k}\gamma_{\bf k^\prime}$
(with $\gamma_{\bf k}=\cos k_x-\cos k_y$)
the pairing interaction in the $d$-wave channel.
(We set the lattice spacing to unity.)
The wave-vector dependence of the gap parameter resulting from Eq. (1)
is purely $d$-wave, $i.e.$, $\Delta_{\bf k}=\Delta \gamma_{\bf k}$, provided
the interaction $V_{{\bf k},{\bf k^\prime}}$ can be 
written in a separable form. This considerably simplifies 
the solution of the above equations. 

We first solve the coupled Eqs.(1) and (2) by the standard Newton's method
with high precision (using a $k$-space mesh up
to $1024\times 1024$ points), 
with the value $t^\prime/t=-0.1$ \cite{notat} to compare
with the inset of Fig. 3 by Soares {\em et al.}. Our numerical 
results for the maximum value of the gap parameter $\Delta (T=0)$
are shown in Fig. 1 as a function of the interaction strength $V/4t$ and for 
different densities. In Fig. 1, the arrows locate the values 
of $V/4t$ below which the gap would vanish (for given density)
according to the numerical results by Soares {\em et al.}. 
We find instead {\em finite} values for the gap even well below 
the critical couplings identified by Soares {\em et al.} 
(and also by den Hertog \cite{Hertog} for the case $t^\prime =0$).
In the weak-coupling regime, our result is that the gap parameter 
decreases exponentially as the coupling is decreased, 
in agreement with the standard BCS result. We obtain the same result
below by analytic calculations.
Regarding instead the chemical potential, our results agree with
those reported in Fig. 3 by Soares et al., as anyway expected by the
smallness of the ratio $\Delta / \mu$ in the weak-coupling region
where our results for $\Delta$ disagree from those by Soares et al.
 
In the low-density regime, where the dispersion is parabolic, 
the $d$-wave BCS equations can be solved {\em analytically} in the 
weak-coupling limit.
The expansion of the dispersion and of the $d$-wave factors 
for small wave vectors leads to
$\xi_{\bf k}\simeq (t+2t^\prime)(k_x^2+k_y^2)-\mu-4t-4t^\prime\equiv
k^2/(2m^*)-\epsilon_F$ and 
$\gamma_{\bf k}\simeq (k_y^2-k_x^2)/2=(k^2/2)\cos 2\phi$, respectively, 
where the polar coordinates $(k,\phi )$ 
have been introduced together with effective mass
$m^*=1/(2(t+2t^\prime))$ and Fermi level 
$\epsilon_F=\mu+4t+4t^\prime$. 
In addition, in the weak-coupling regime it is convenient to limit 
the integral over the energy variable $\epsilon$ to a small window 
$\mid \epsilon - \epsilon_F \mid < \omega_0/2$
about the Fermi level, 
where $\omega_0$ is a cutoff such that $\omega_0\gg \Delta$.
The integral outside this energy window gives sub-leading 
contributions that can only be evaluated numerically since the full
wave-vector dependence of $\xi_{\bf k}$ and $\gamma_{\bf k}$ 
ought to be retained in this case. 
Note that, in the weak-coupling regime, the Fermi level coincides
with the non-interacting value $\epsilon_F=k_F^2/(2m^*)=\pi n /m^*$
and Eqs. (1) and (2) can be decoupled. 
After these manipulations,
we are led to the following expression for 
the $d$-wave gap equation at $T=0$: 

\begin{equation}
\label{lowd}
1=\frac{2g_d}{\pi}\int_{-\omega_0/2}^{\omega_0/2}d\xi
\int_{0}^{\frac{\pi}{2}}d\theta \frac{\cos^2\theta}
{\sqrt{\xi^2+\Delta_d^2\cos^2\theta}}
\end{equation}
where $g_d=Vk_F^4m^*/(16\pi)$, $\Delta_d=\Delta k_F^2/2$,
$\theta=2\phi$, and $\xi = \epsilon - \epsilon_F$. 
Equation (3) has been written in a form which maps exactly into
Eq. (A2) of Ref.\onlinecite{Musaelian}. The analytic solution for the 
gap equation in the weak-coupling limit reported in Ref.\onlinecite{Musaelian}
clearly shows that the gap is {\em always finite}, no matter how weak the
effective attraction $g_d$ is. Since this point is crucial to the present
discussion, we provide here some details of the derivation of the analytic
result given by Eq. (A2) of Ref.\onlinecite{Musaelian}.
The integral over the angular variable in Eq. (3) can be performed 
exactly using known results for the elliptic integrals \cite{Marini}. 
One obtains:
\begin{eqnarray}
\int_{0}^{\frac{\pi}{2}}d\theta \frac{\cos^2\theta}
{\sqrt{\xi^2+\Delta_d^2\cos^2\theta}}=
\frac{1}{\sqrt{\xi^2+\Delta_d^2}}
\int_{0}^{\frac{\pi}{2}}d\theta \frac{\cos^2\theta}
{\sqrt{1-\kappa^2\sin^2\theta}}\nonumber\\
=\frac{1}{\sqrt{\xi^2+\Delta_d^2}}\left[
\frac{1}{\kappa^2}E(\frac{\pi}{2},\kappa)-
\frac{(1-\kappa^2)}{\kappa^2}F(\frac{\pi}{2},\kappa)
\right]
\end{eqnarray}
where $\kappa^2=\Delta_d^2/(\xi^2+\Delta_d^2)$, and  
$F(\frac{\pi}{2},\kappa)$ and $E(\frac{\pi}{2},\kappa)$ are elliptic 
integrals of the first and second kind, respectively.
Performing at this point the integral over the variable $\xi$,
the expansion of the elliptic functions 
$E(\frac{\pi}{2},\kappa)$ and $F(\frac{\pi}{2},\kappa)$
up to second-order in $\kappa$ produces a log contribution in 
$\omega_0/\Delta_d$, leading to the usual
BCS exponential dependence of the gap parameter on coupling. 
The prefactor of this exponential is obtained by considering
the sum of the remaining terms of the above expansion, 
resulting in finite integrals over the variable $\xi$ as 
$\omega_0/\Delta_d\rightarrow \infty$.
These terms can be summed exactly by exploiting the analytic result
\begin{equation}
\int_{0}^{1}d\kappa \frac{1}
{\kappa \sqrt{1-\kappa^2}} \left[
\frac{1}{\kappa^2}E(\frac{\pi}{2},\kappa)-
\frac{(1-\kappa^2)}{\kappa^2}F(\frac{\pi}{2},\kappa)-\frac{\pi}{4}
\right]=\frac{\pi}{4}\ln\left(\frac{2}{\sqrt{e}}\right).
\end{equation}
The final expression for the gap parameter $\Delta_d$ obtained in this way
is:
\begin{equation}
\label{res}
\Delta_d = \frac{2\omega_0}{\sqrt{e}} \; \exp (-1/g_d)\; .
\end{equation}
Note that the coupling constant 
$g_d=\pi n^2 V m^*/4$ entering the exponential in Eq. (6) depends on the
squared density, leading to a marked exponential suppression
of the gap parameter in the weak-coupling regime as the low-density limit
$(n\rightarrow 0)$ is approached. We have verified that our numerical 
results for the gap parameter reproduce this  
dependence on coupling  strength $V$ and density $n$.
We emphasize, however, that only the exponential dependence in Eq. (6)
can be compared with the numerical results, since the sub-leading
contributions to the energy integral in Eq. (3) outside the 
$\omega_0$-window (which are not included in the present calculation)
are expected to eventually
modify the prefactor of Eq. (6) in such a way to eliminate 
this cutoff completely (the energy scale $\omega_0$ is  
absent in our starting model and has been introduced only 
for convenience in the analytic calculation). 

Summarizing, we have found that a non-pairing region is
completely absent in the
phase diagram of the attractive Hubbard model for $d$-wave pairing
describing the BCS-BE crossover. The gap parameter at $T=0$ is, in fact,
finite for any non-zero density and coupling, even in the weak-coupling 
regime, contrary to the results of Soares {\em et al.} \cite{Soares}
and den Hertog \cite{Hertog}.

{\em Acknowledgements.} We are indebted to David Neilson 
for a critical reading of the manuscript.
A. Perali gratefully acknowledges financial support 
from the Italian INFM under contract PAIS Crossover No. 269.

\begin{figure}
\centerline{\psfig{figure=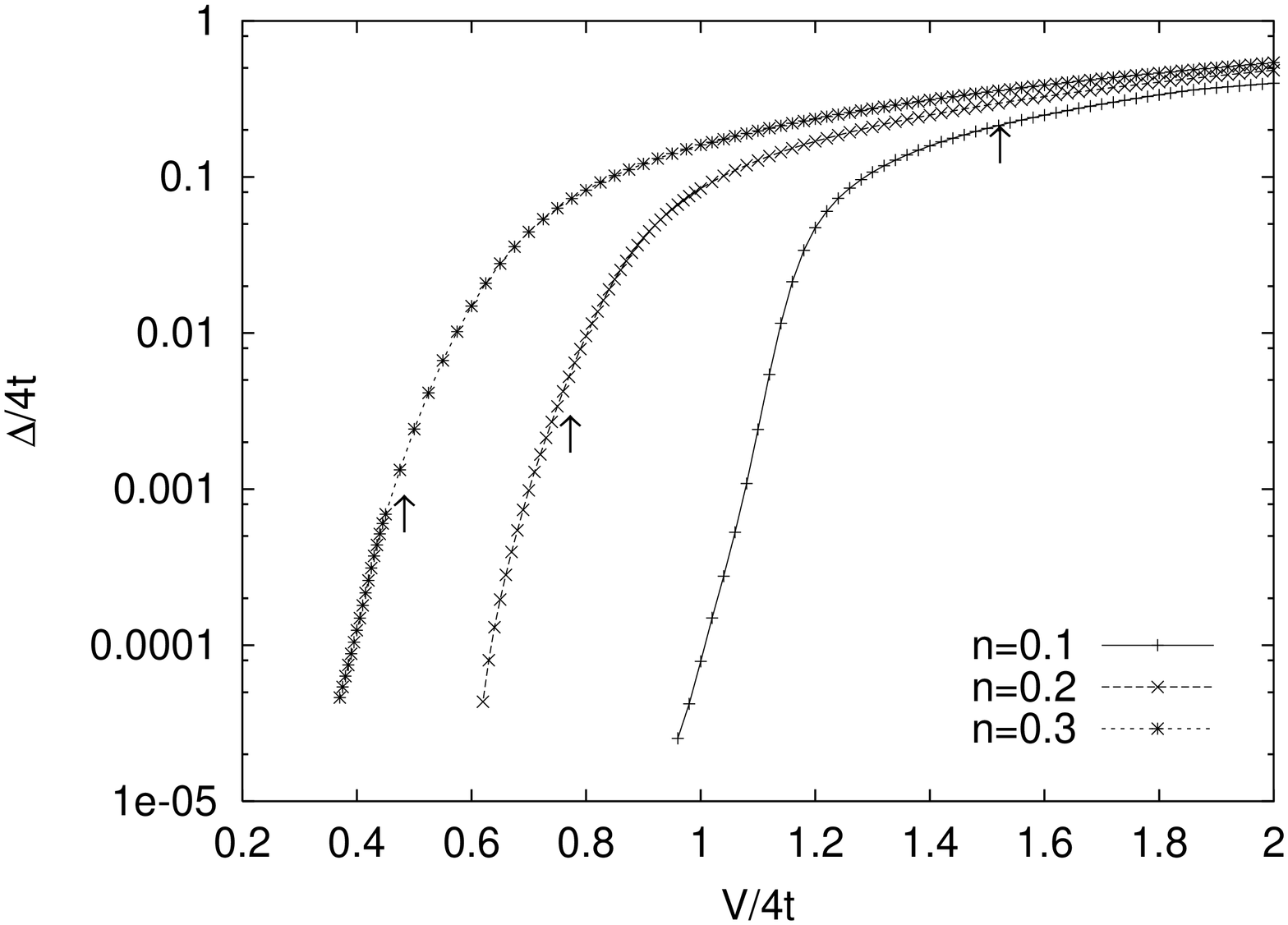,width=7cm,angle=-90}}
\caption{Maximum value of the $d$-wave gap parameter $\Delta$ at $T=0$ as 
function of the interaction strength $V$ 
for different densities [$n=0.1$ (plus), $n=0.2$ (cross), $n=0.3$ (star)].
Energies are normalized with respect to the half-bandwidth $4t$.
The arrows locate the values of $V/4t$ below which Ref.[1]
claims that $\Delta =0$.}
\label{f1}
\end{figure}

\end{document}